\documentclass[fleqn,twoside]{article}
\usepackage{espcrc2}

\usepackage{graphicx}
\usepackage[figuresright]{rotating}
\usepackage{epsfig}
\usepackage{url}

\newcommand{\AmS}{{\protect\the\textfont2
  A\kern-.1667em\lower.5ex\hbox{M}\kern-.125emS}}
\title{ TAUOLA for simulation of tau decay and production:
perspectives for precision low energy and LHC
applications }

\author{ Z. W\c{a}s\address[MCSD]{Institute of Nuclear Physics, Polish Academy of Sciences,\\
         ul. Radzikowskiego 152, 31-342 Cracow, Poland}
        \thanks{  Supported in part  by the Polish Government grant
 N202 06434 (2008-2011). }}

\begin{document}

\begin{abstract}
The status of Monte Carlo system for the simulation of $\tau$-lepton production and decay in 
high-energy accelerator experiments is reviewed. 
Since previous $\tau$-lepton conference in 2008 some practical modifications 
have been  introduced:
 (i) For the {\tt TAUOLA} Monte Carlo generator of $\tau$-lepton decays,  
automated and simultaneous use of  many versions of 
form-factors for the calculation of optional weights  for fits was developped
and checked to work in Belle and BaBar software environment. 
Work on alternative paramterizations of hadronic decays is advanced.
(ii)  
the  {\tt TAUOLA universal interface} based on {\tt HepMC} (the {\tt C++} 
event record) is now public. A similar interface for {\tt PHOTOS} is now 
also public. 
(iii) Extension  of {\tt PHOTOS} Monte Carlo for QED bremsstrahlung 
in  decays featuring kernels based on complete first order matrix element are 
gradually becoming widely available thanks to properites of the new,
{\tt HepMC} based interface. (iv) Tests of the programs systematized 
with  the help of {\tt MC-TESTER} are now available for {\tt FORTRAN} and C++ 
users.

Presented here results illustrate  the status of the projects performed in 
collaboration with  Nadia Davidson, Piotr Golonka,
Gizo Nanava, Tomasz Przedzi\'nski, Olga Shekhovtsova, 
El\.zbieta Richter-W\c{}as, Pablo Roig, Qingjun Xu and others.

\vspace{1mm}
\centerline{ \it Presented at International workshop on Tau Lepton Physics, TAU10 Manchester GB, September, 2010}
\vspace{1pc}
\centerline{preprint \hskip 1 cm {\bf  IFJPAN-IV-2011-1}, \hskip 1 cm January 2011 \hskip 7 cm}
\vspace{1pc}
\end{abstract}

\maketitle

\setcounter{footnote}{0}
 
\section{Introduction}

The {\tt TAUOLA} package
\cite{Jadach:1990mz,Jezabek:1991qp,Jadach:1993hs,Golonka:2003xt}  for the simulation 
of $\tau$-lepton decays and  
{\tt PHOTOS} \cite{Barberio:1990ms,Barberio:1994qi,Golonka:2005pn} for the simulation of QED radiative corrections
in decays, are computing
projects with a rather long history. Written and maintained by 
well-defined (main) authors, they nonetheless migrated into a wide range
of applications where they became ingredients of 
complicated simulation chains. As a consequence, a large number of
different versions are presently in use. Those modifications, especially in case of
{\tt TAUOLA}, are   valuable from the physics point of view, often did not found the place in the distributed versions of
the program.  Even if 
from the algorithmic point of view, versions
differ only in  details, they incorporate many specific results from distinct
$\tau$-lepton measurements or phenomenological projects. 
Such versions were mainly maintained (and will remain so) 
by the experiments taking precision data on $\tau$ leptons. 
Interesting from the physics point of view changes are expected to be 
developped in {\tt FORTRAN}.
That is why, significant part of the 
{\tt TAUOLA} should remain in {\tt FORTRAN} for the forthcoming several years.

However
 many new applications were developed  recently,  often requiring
a program interface to other packages  (e.g. generating events for LHC, LC, 
Belle or BaBar physics processes). 
Fortunately, co-existence with C++ is not a problem, at least 
not from 
the software point of view.

The programs structure,
prepared for  the convenience of {\tt FORTRAN} users,  was presented during previous 
$\tau$ conferences, 
and we will not repeat it here. 
This time, let us concentrate on new interfaces 
for applications based on {\tt HepMC} \cite{Dobbs:2001ck} event record.
We will also report on progress in implementation (re-implementation) of 
techniques useful  for fits.
Analyses of high precision,
high statistic  data from Belle and BaBar are expected to progress from these
solutions.

Our presentation is organized as follows: 
Section 2 is devoted to new interfaces of {\tt TAUOLA} and {\tt PHOTOS} to 
{\tt HepMC} applications of C++, where genuine weak and transverse spin effects can be taken into account as well.
Section 3  is devoted to the discussion  of optional 
weights in {\tt TAUOLA} and their use for fits to data at 
the level of comparison with raw data. Progress on the work on new
currents for hardonic decays which can be confronted with (tuned to) 
data using such optional weights enabling simultaneous control of all 
experimental effects is mentioned too.
In section 4 we present some new results for   {\tt PHOTOS} Monte Carlo for 
radiative corrections in decays. 
Section 5 is devoted to {\tt MC-TESTER}, the program which can be used for 
semi-automatic comparisons of simulation samples originating from
different programs.

Because of the limited space of the contribution, 
and sizable amount of other physically interesting 
results, some of them will be excluded from conference 
proceedings. They will find place in 
future works, possibly with collaborators mentioned in the Abstract.
For these works,  the present paper may serve as an announcement.

\section{  {\tt TAUOLA universal interface} and {\tt PHOTOS} interface in C++}

\begin{figure}
\begin{center}
\setlength{\unitlength}{0.5 mm}
\begin{picture}(35,80)
\put( -65,-45){\makebox(0,0)[lb]{\epsfig{file=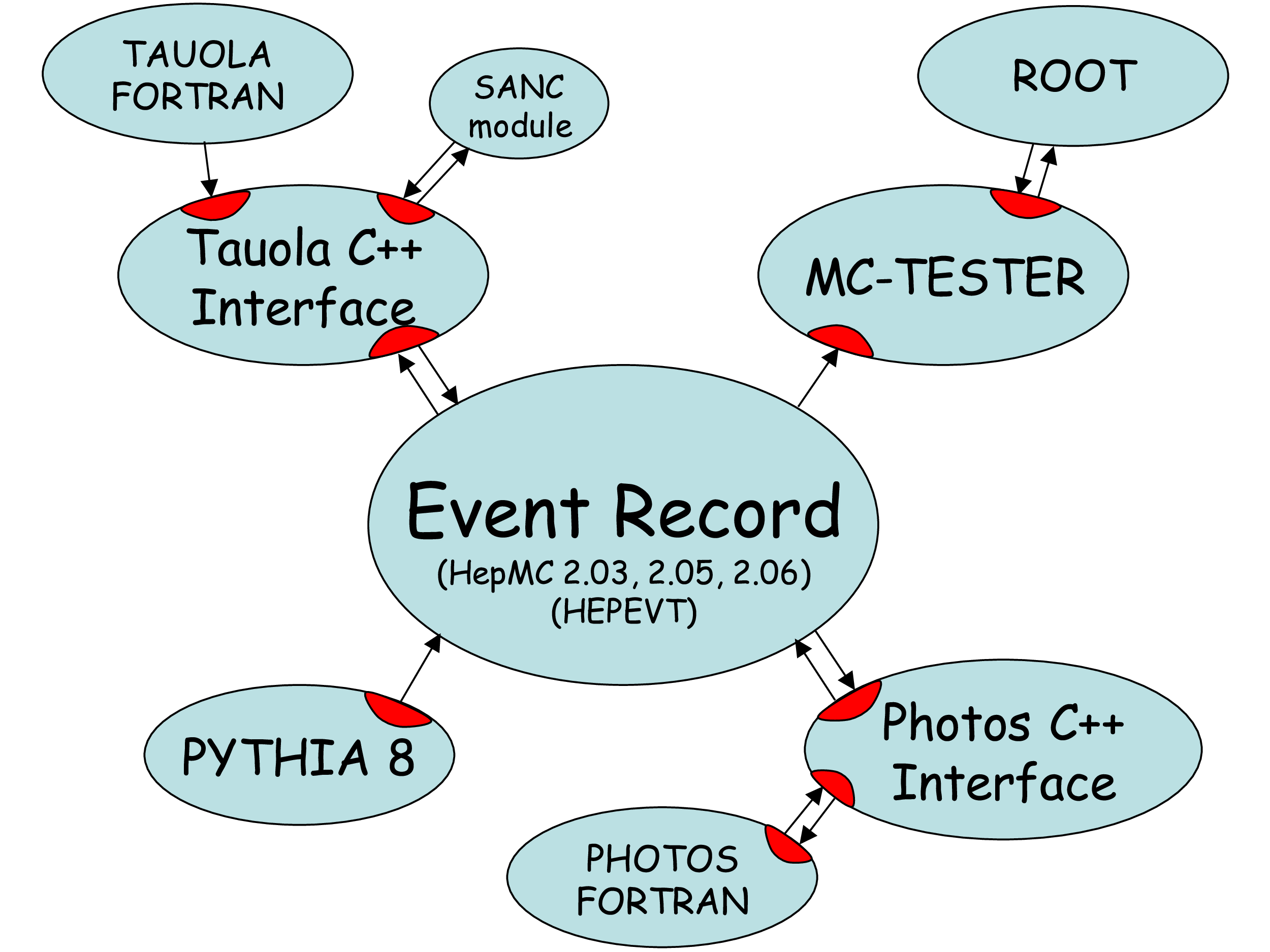,width=70mm,height=65mm}}}
\end{picture}
\end{center}
\vskip 1.5 cm
\caption{\small \it  Scheme of Monte Carlo simulation system with communication based on event record.
 } \label{Relations}
\end{figure}

\begin{figure}[!ht]
\setlength{\unitlength}{0.1mm}
\begin{picture}(800,1250)
\put(5, 25){\makebox(0,0)[lb]{\epsfig{file=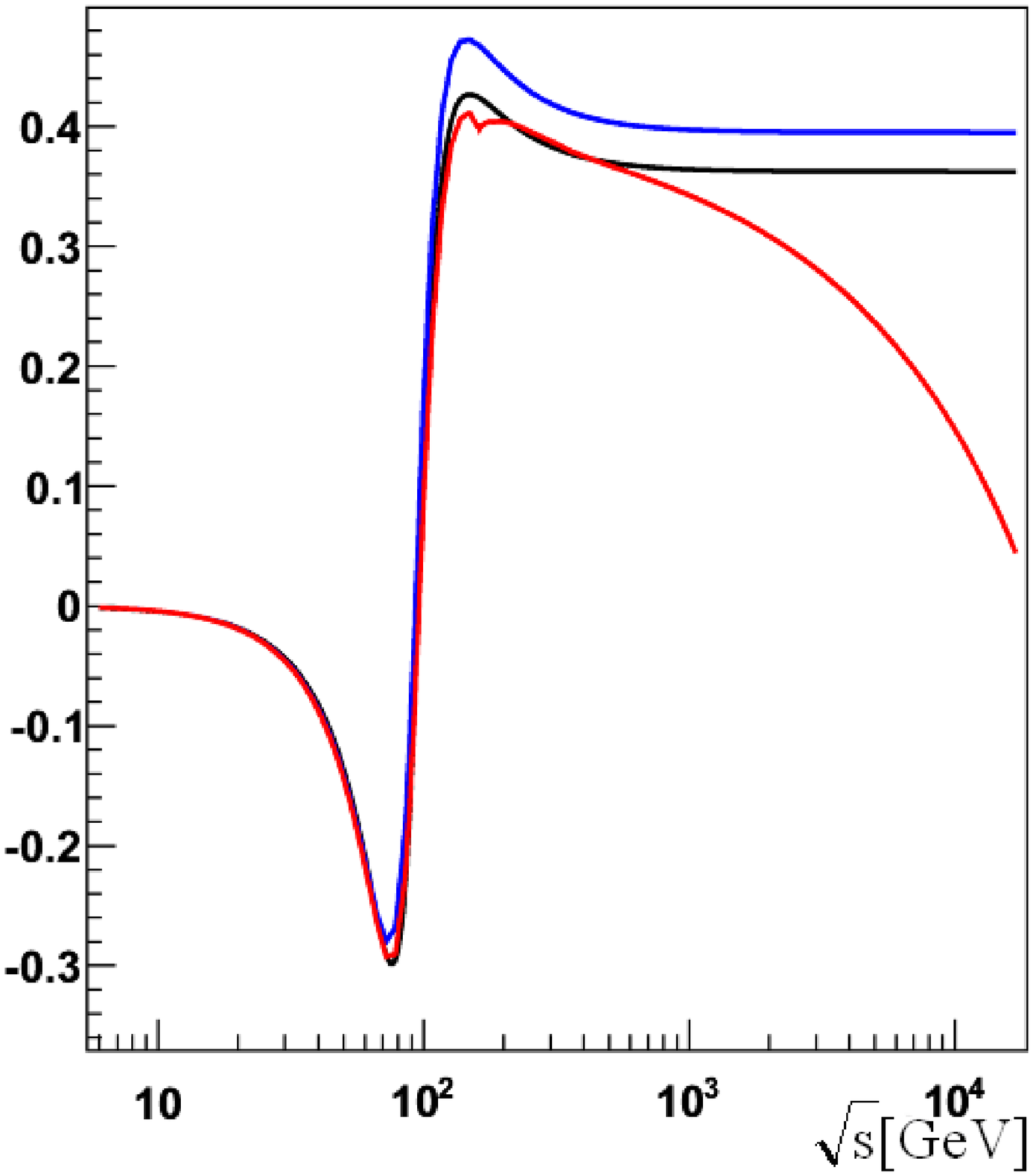,width=75mm,height=60mm}}}
\put( 55,-15){\makebox(0,0)[lb]{\bf a}}
\put(5,650){\makebox(0,0)[lb]{\epsfig{file=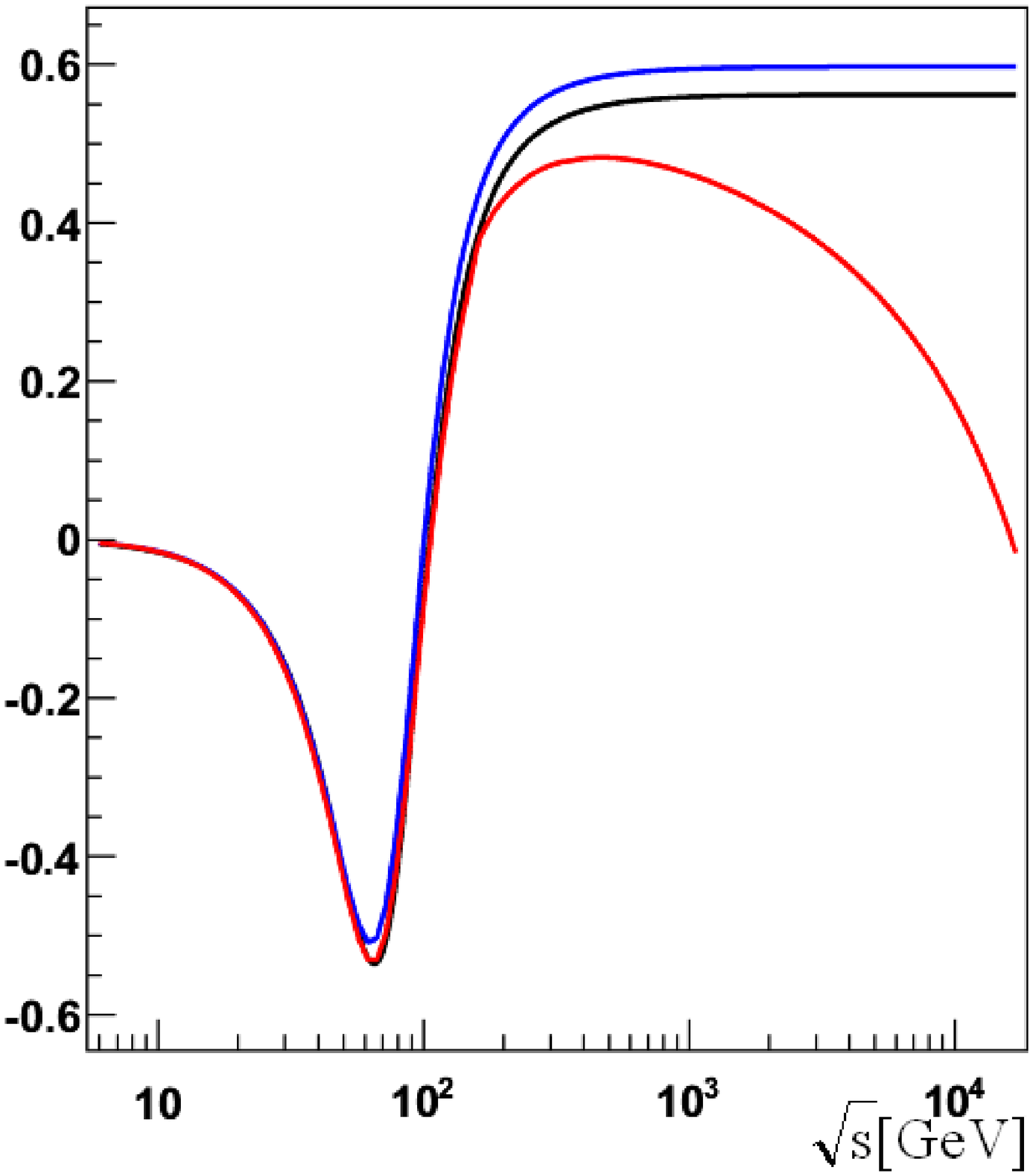,width=75mm,height=60mm}}}
\put( 55,650){\makebox(0,0)[lb]{\bf b}}
\end{picture}
\caption{ \small \it Polarization of $\tau$ leptons produced from up quarks
  (Fig. (a)) and down quarks (Fig. (b)) at $\cos\theta=-0.2$ 
($q\bar q \to \tau^+\tau^-$).  For the red
  line genuine weak corrections are taken into account, the black represent Standard Born (default in the interface). The blue line is Born according to
  alpha scheme of ~\cite{Andonov:2008ga,Andonov:2004hi}.   The small bump on the red
  line on Fig. (a) is due to the WW threshold.  }\label{fig:polDown}
\end{figure}

In the development of packages such as {\tt TAUOLA} or {\tt PHOTOS}, questions 
of tests and appriopriate relations to users' applications are essential for 
their 
usefulness. In fact, user applications may be much larger in size and 
human effort, than the programs discussed here. 
Good example of such `user applications' are complete environments to simulate 
physics process and control detector response at the same time. 
Distributions of final state particles distributions are not always of direct interest. 
Often properties of intermediate states such as spin state of $\tau$-lepton, coupling constants or masses of intermedate heavy particles are of prime interest.
As a consequence it is useful that such intermediate state properties are
under direct control of the experimental user and can be manipulated 
to understand detector responses.
Our programs  worked well   with {\tt FORTRAN} applications  where {\tt HEPEVT} event record 
was used.  Now, for the  {\tt C++}  {\tt HepMC} \cite{Dobbs:2001ck} case, 
interfaces had to be re-written, both for 
{\tt TAUOLA} \cite{Davidson:2010rw} and {\tt PHOTOS} \cite{Davidson:2010ew}.  
The interfaces are gradually enriched and for example  for the {\tt PHOTOS} 
improved kernel,  is already available 
for 
the general use, as explained in ref.~\cite{Golonka:2006tw}. 
 A complete (not  longitudinal only) spin correlations 
 are available 
for $Z/\gamma^*$ decay in {\tt TAUOLA} interface now. 
They are based on electroweak corrections taken from 
refs.~\cite{Andonov:2008ga,Andonov:2004hi}.
See Fig.~\ref{Relations} for the scheme 
of programs communications. The size of the electroweak effects on longitudinal
$\tau$ polarization is shown in Fig.~\ref{fig:polDown}.
Such modular organization opens ways for efficient algorithms to understand 
detector systematics, but at the same time responsability to control software
precision must be shared by the user.
For that purpose 
automatization of tests
{\tt MC-TESTER} was prepared~\cite{Golonka:2002rz}.
New functionalities
were introduced into the testing package \cite{Davidson:2008ma}. In particular, it works now with the
{\tt HepMC} event record, the  standard of {\tt C++} programs and the 
spectrum of available tests is enriched. For details see Section 5.

\section{ Optional weights for  {\tt TAUOLA} Monte Carlo} 
Physics of $\tau$ lepton decays requires sophisticated strategies for the
confrontation of phenomenological models with experimental data. On one hand 
high statistics experimental samples are collected, and the obtained precision is 
high, on the other hand, there is a significant cross-contamination between distinct
$\tau$ decay channels. Starting from  a certain precision  level all channels 
need to be analyzed simultaneously. Change of parametrization for one channel 
contribution to the background may be important for the fit of another
 one. This situation leads to a complex configuation where a multitude of parameters (and models)
needs to be simultaneously confronted with a multitude of observables.
One has to keep in mind that the models used to obtain distributions in
 the fits may require refinements or even substantial rebuilds as a consequence
of comparison with the data. The topic was covered in detail in the $\tau$ Section of Ref.~\cite{Actis:2010gg} earlier this year.

\begin{figure}[h!]
\centering
\includegraphics[scale=0.32]{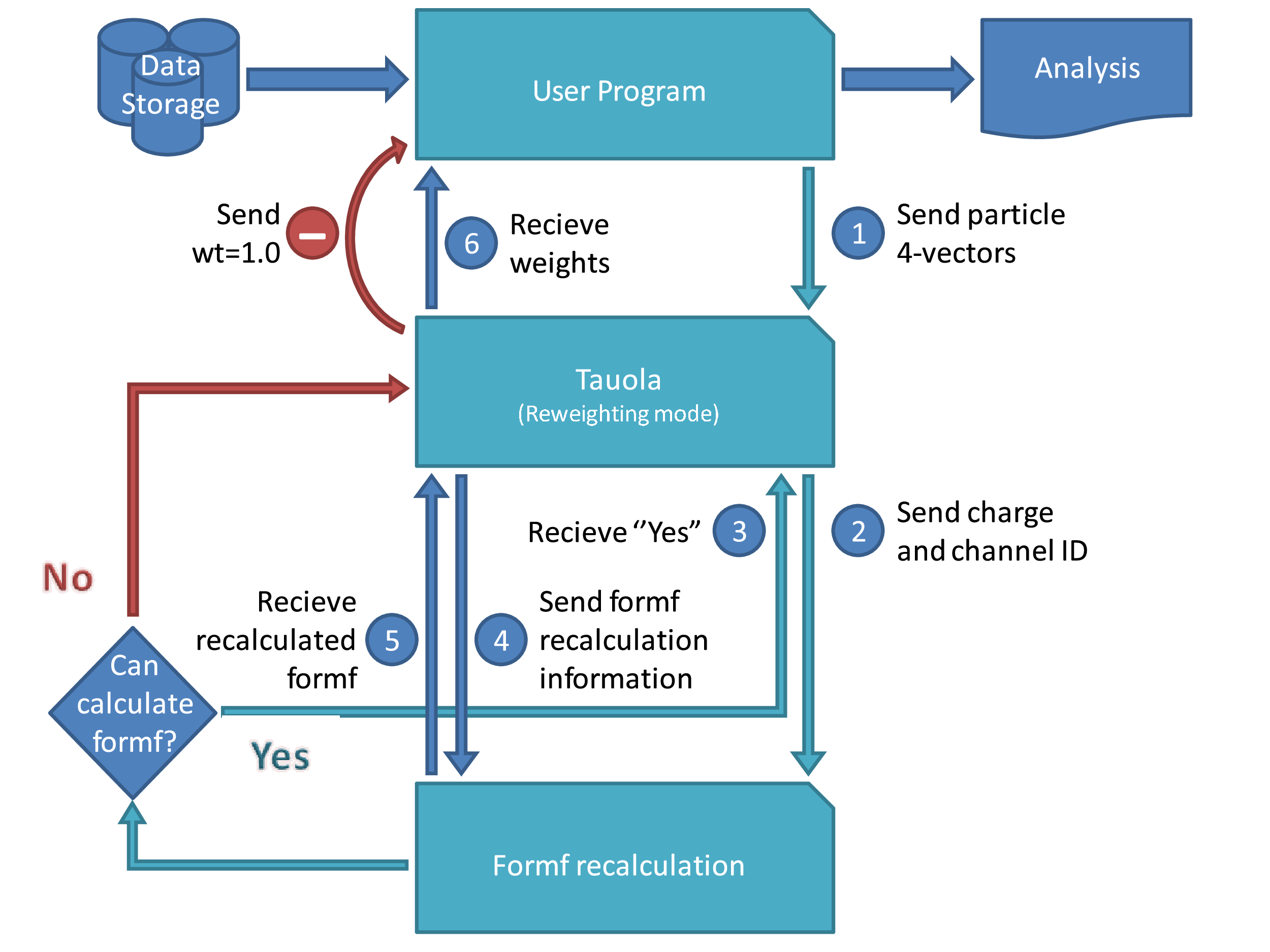}
\label{fig:doxy}
\caption{ \it
Flow chart for fifo communication and reweigting {\tt TAUOLA} matrix elements, 
verified to be compatible with Belle and BaBar software.
} \label{Flow}
\end{figure}

From the statistical point of view it is best to resolve such system in one 
automated step using  for example a method 
such as \cite{tmva,Hocker:2007ht}. 
This can be of course very dangerous from 
the point of view of systematic error control. But we will not elaborate on this
point any further. From the technical point
 it is necessary, to calculate for each generated event (for each
present in it: decay of $\tau^+$ and $\tau^-$ separately) alternative weights; the ratios
of matrix element squared obtained with new currents  provided by the user,
and the one actually used in generation. Then  the vector of weights can be obtained
and used in fits.
We have checked that such a solution not only can be easily installed into
{\tt TAUOLA} as a stand-alone generator but it can also be incorporated into 
the simulation frameworks of Belle and BaBar collaborations rather easily. 
The scheme as given in Fig.~\ref{Flow} was shown to work in  real conditions.

At the present step of our work no new models of hadronic currents are  
prepared. Work on  currents based on refs. \cite{Roig:2008xt,Dumm:2009kj,Dumm:2009va}
is taking shape mainly thanks to efforts by Olga Shekhovtsova. 
A dedicated patch is being prepared, to be later installed into users software environments.

\section{{\tt PHOTOS} Monte Carlo for bremsstrahlung and its systematic uncertainties}
\def\CCol{{\tt SANC}}
Thanks to exponentation properties and factorization, the bulk of the final state 
QED bremsstrahlung can be described in a universal way.   
Howevers, the 
kinematical configurations caused by QED bremsstrahlung are affecting 
in an  important way
signal/background separation. It may affect selection criteria and background 
contaminations in quite complex and unexpected ways. 
In many applications, not only in $\tau$ decays,
such bremsstrahlung corrections are 
generated with the help of the  {\tt PHOTOS} Monte Carlo. That is why it is of importance to
review the precision of this program as documented in 
Refs.~\cite{Barberio:1990ms,Barberio:1994qi,Golonka:2005pn}. 
For the C++ applications, the first version of the program is also available now, 
it is documented in Ref.~\cite{Davidson:2010ew}. 

In C++ applications, the complete first order matrix element  for the 
 two body decays of the $Z$ into pair of charged leptons  
 \cite{Golonka:2006tw} is now available.
Kernels with complete matrix elements, for the decays of 
scalar $B$ mesons  into a pair of scalars  \cite{Nanava:2006vv},  
 for  $W^\pm \to l^\pm \nu \gamma$ 
and  for $\gamma^* \to \pi^+\pi^-$ \cite{Nanava:2009vg} are available for tests 
or specially oriented decay particles frames.
It will be rather easy now to  
integrate NLO kernels  into the main version of the program,
 because of better control of 
decay particle rest frame than in the {\tt FORTRAN} interface.

In all of these cases the universal kernel of {\tt PHOTOS} is replaced with the 
one matching exact first order matrix element. In this way terms necessary 
for NLO/NLL precision 
level are implemented\footnote{Note that here  the LL (NLL) denote  
 collinear logarithms (or in case of differential 
predictions terms integrating into such logarithms).
 The logarithms of soft singularities are taken into 
account to all orders. This is resulting from mechanisms of exlusive 
exponentiation \cite{Jadach:2000ir} of QED. 
The algorithm used in {\tt PHOTOS} Monte Carlo is compatible with exclusive 
exponentiation. Note that our
 LL/NLL precision level would even read  as  respectively   NLL/NNNLL
 level in 
some naming conventions of QCD.
}. 
A discussion relevant for control of program systematic uncetainty in $\tau \to \pi \nu$ decay can be found in 
Ref.~\cite{Guo:2010ny}. 

The algorithm covers the full multiphoton 
phase-space and becomes  exact in the soft limit. 
This is rather unusual for  NLL compatible algorithms. One should not forget 
that {\tt PHOTOS} generates weight-one events too, and does not require any 
phase space ordering. There is a full phase space overlap between the one 
where  hard 
matrix element is used and the one for iterative photon emissions.
All interference effects (between consecutive emissions and emissions from 
distinct charged lines) are implemented with the help of internal weights.

The results of all tests of {\tt PHOTOS} with a NLO kernel confirm
 sub-permille precision level.
This is very encouraging, and points to the possible extension of the 
approach outside of  QED (scalar QED). In particular to the domain of 
QCD or if phenomenological lagrangians for interactions of photons need
 to be applied. For that work  to be completed spin amplitudes need 
to be further studied. Let us point to Ref.~\cite{vanHameren:2008dy} 
as an example.

\begin{figure}[!ht]
\setlength{\unitlength}{0.1mm}
\begin{picture}(800,1250)
\put(5, 25){\makebox(0,0)[lb]{\epsfig{file=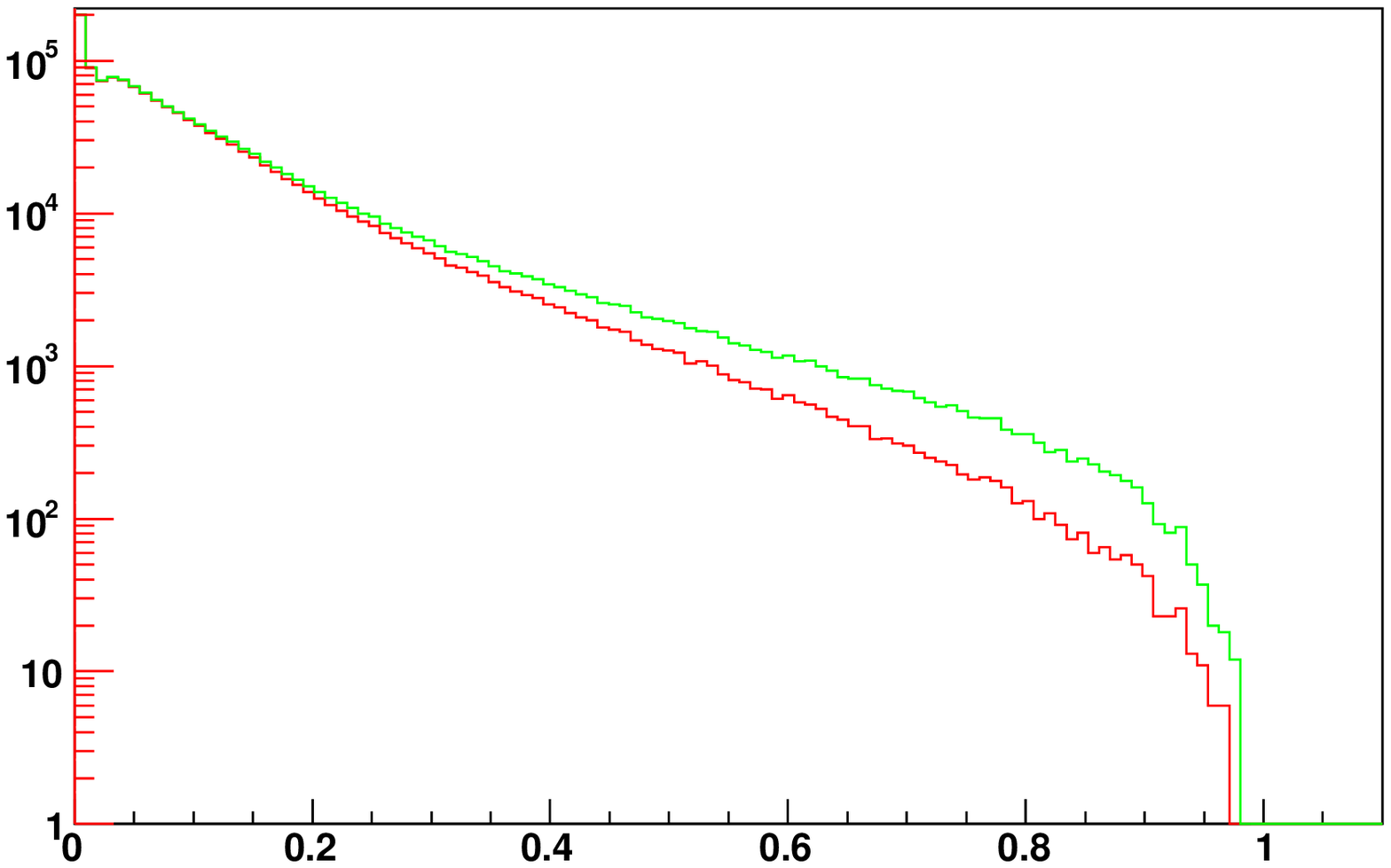,width=75mm,height=60mm}}}
\put( 55,-15){\makebox(0,0)[lb]{\bf a}}
\put(5,650){\makebox(0,0)[lb]{\epsfig{file=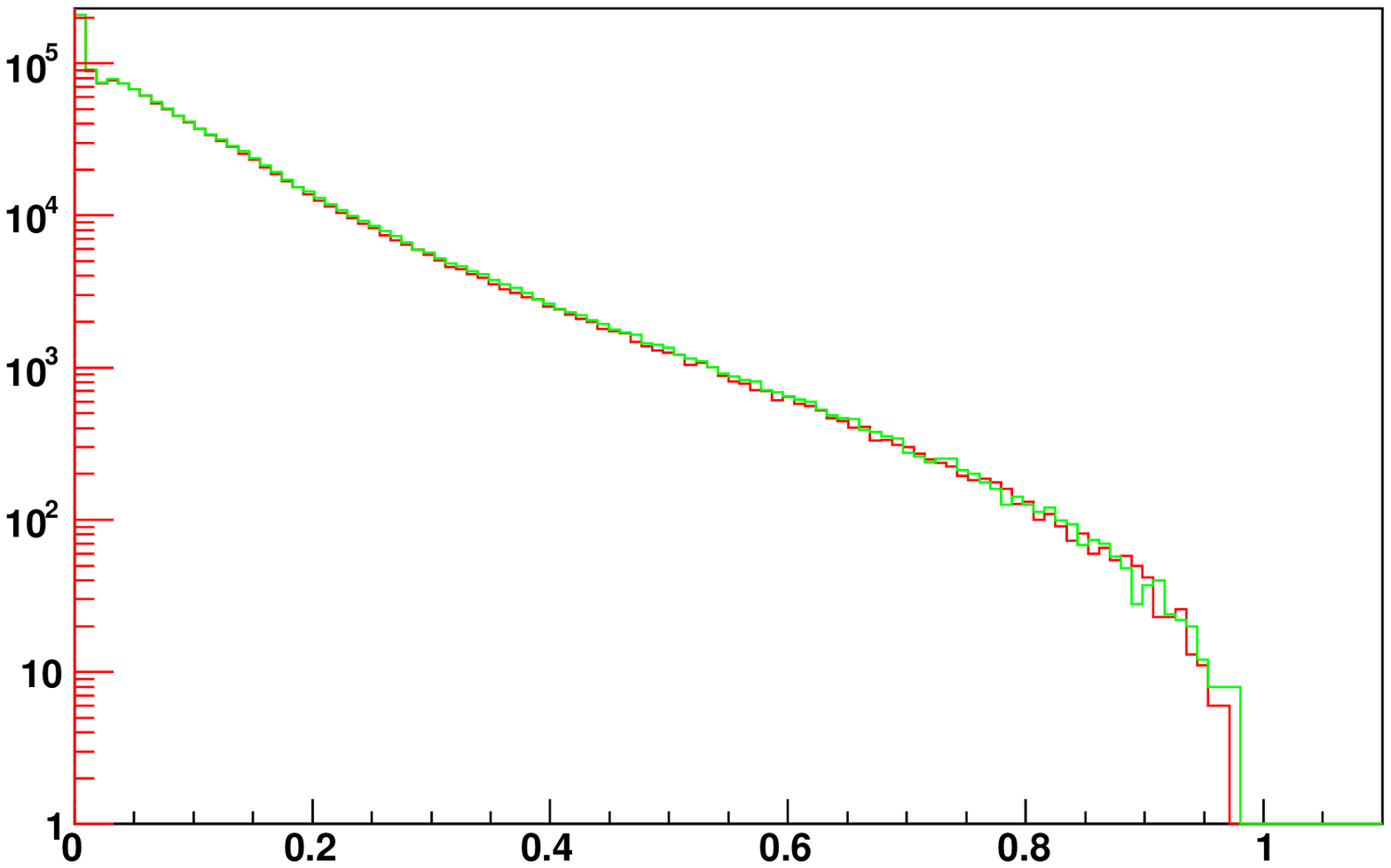,width=75mm,height=60mm}}}
\put( 55,650){\makebox(0,0)[lb]{\bf b}}
\end{picture}
\caption{ \it The spectrum of the $\gamma \gamma$ invariant mass in $Z \to \mu^+\mu^- n\gamma$
decay. Events with two hard photons, both of energy above 1 GeV in the $Z$ rest 
frame are taken and the invariant mass of the photon pair, normalized to Z mass
is shown: for CEEX2 and CEEX1 (case a), and CEEX2 and {\tt PHOTOS}
(case b). The prediction from {\tt PHOTOS} is clearly superior for applications 
aiming at
simulations for Higgs boson backgrounds than CEEX1. In the case of solution 
based on YFS 
exponentiation \cite{Jadach:1999vf}, the second order matrix element must be taken into account. 
Fig.~{\bf b} was obtained with the help of 
 {\tt examples/testing/Zmumu} {\tt PHOTOS} demonstration example (as
documented  in Ref.~\cite{Davidson:2010ew}). 
  \label{fig:gamgam}
}
\end{figure}

Kernels necessary for NLO are only available at present as options for tests 
only.
 They can be uploaded  from the {\tt PHOTOS} web page
\cite{Photos_tests}. In the C++ applications they will be gradually installed
into the main version of the program. The first step was the case of 
$Z/\gamma^*$ decay and is already completed now. This opens the way to discuss 
systematic errors in user defined acceptance regions. This will be an important 
supplement to tests-comparisons
 with other programs such as KKMC \cite{Jadach:1999vf},
 see Fig.~\ref{fig:gamgam}.

\section{{\tt MC-TESTER} and user defined tests.}
Our work on {\tt MC-TESTER}~\cite{Davidson:2008ma} reached maturity. 
As in the past, the program
main purpose remain benchmarking the decay part of different Monte Carlo chains. 
Generated events have to be  stored in event records: be it of {\tt FORTRAN}
 or C++.  Default distributions consist of all 
possible invariant masses which are automatically generated and stored for 
each found decay channel of the particle under test. 
Then, at the analysis step,  information from a pair of such runs may
 be compared and represented in the form of tables and plots. 
At present, users macros can be easily installed, in particular all 
demo distributions given in papers on C++ interfaces for  
{\tt TAUOLA} \cite{Davidson:2010rw} and
{\tt PHOTOS} \cite{Davidson:2010ew} can be obtained in that way. 
Set-up's  for benchmarking the interfaces, such as interface
 between $\tau$-lepton
production and decay, including QED bremsstrahlung effects 
can be prepared in that way.

The updated version of {\tt MC-TESTER} was found useful
for {\tt FORTRAN} ~\cite{Golonka:2005pn,Golonka:2006tw}
and for C++ \cite{Davidson:2010ew}  examples 
 where spurious information (on soft photons)
was removed.

Finally, let us mention that the program is available 
through the Grid Project LCG/Genser web page, see Ref.~\cite{Kirsanov:2008zz} for details. This is the case for {\tt TAUOLA} C++ and for 
{\tt PHOTOS}  C++ in the near future as well. 
The {\tt FORTRAN} predecessors have already been  
available for some time.

\section{Summary and future possibilities}

The status of the computer programs for the decay of $\tau$ leptons
{\tt TAUOLA}
 and
associated projects {\tt TAUOLA universal interface } and {\tt MC-TESTER}  
was reviewed.
The high-precision version of  {\tt PHOTOS} for radiative corrections in 
decays, was 
presented also. All these programs are available now for C++ applications 
thanks to the {\tt HepMC} interfaces.
 
New results for  {\tt PHOTOS}   were mentioned.
  For the $Z$ decay channel the complete next-to-leading collinear logarithms
effects can now be  simulated in C++ applications. 
However, in  most cases these 
effects are not important, leaving the standard  version 
of {\tt PHOTOS} sufficient.
The example is  important for the general case, it helps 
to better understand questions related to matching hard emission matrix 
elements with parton showers without the necessity to introduce any boundaries
within the phase space.
 Thanks to this work  the path for  fits to the data of 
electromagnetic form-factors  
 is opened.
 The effects of these  form-factors may be significantly larger 
and physically more interesting than  complete next-to-leading order
effects of QED or scalar QED.

 The presentation of the {\tt TAUOLA} general-purpose interface
in {\tt C++} was given. It is now more refined
than the {\tt FORTRAN} predecessor. Electroweak corrections can be used 
in calculation of complete spin correlations in $Z/\gamma^*$ mediated 
processes.  

Distinct versions of the hadronic currents for {\tt TAUOLA} library 
are available for the FORTRAN distribution.  Work on  new ones 
and on refinements of old strategies of fits using alternative weights and 
projection operators is progressing.

The new version of {\tt MC-TESTER} is stable now.
It not only works with {\tt HepMC}  of {\tt C++} but enables
user defined tests in experiments' software environments.

\vskip 1 mm
\centerline{ \bf Acknowledgements}
\vskip 1 mm

Discussions   
with  members of the Belle and BaBar collaborations 
are  acknowledged. Exchange of e-mails and direct discussions 
with  S. Banerjee, S. Eidelman, H. Hayashii, K. Inami,  A. Korchin, J. H. K\"uhn   and 
M. Roney was a valuable  input to present and future steps in 
projects development.

\providecommand{\href}[2]{#2}
\end{document}